\documentclass{article}
\usepackage{authblk}

\usepackage{amsfonts}
\usepackage[cmex10]{amsmath}
\usepackage{amssymb}
\usepackage{bbm}

\usepackage{graphicx}

\usepackage{multirow}
\usepackage{ltxgrid}
\usepackage{subfig}

\usepackage[margin=25truemm]{geometry}

\usepackage{url} 

\author[1,*]{\small Taketo Akama}
\author[1]{\small Hiroaki Kitano}
\author[2]{\small Katsuhiro Takematsu}
\author[2]{\small  Yasushi Miyajima}
\author[1]{\small Natalia Polouliakh}

\affil[1]{\footnotesize Sony Computer Science Laboratories, Inc, Tokyo, Japan}
\affil[2]{\footnotesize Koozyt, Inc, Tokyo, Japan}
\affil[*]{Corresponding author: Taketo Akama, taketo.akama@sony.com}

\date{}
\title{Self-supervised Auxiliary Loss for Metric Learning in Music Similarity-based Retrieval and Auto-tagging}

\begin{document}
  \maketitle
\begin{abstract}
In the realm of music information retrieval, similarity-based retrieval and auto-tagging serve as essential components. Given the limitations and non-scalability of human supervision signals, it becomes crucial for models to learn from alternative sources to enhance their performance.
Self-supervised learning, which exclusively relies on learning signals derived from music audio data, has demonstrated its efficacy in the context of auto-tagging. In this study, we propose a model that builds on the self-supervised learning approach to address the similarity-based retrieval challenge by introducing our method of metric learning with a self-supervised auxiliary loss. Furthermore, diverging from conventional self-supervised learning methodologies, we discovered the advantages of concurrently training the model with both self-supervision and supervision signals, without freezing pre-trained models. We also found that refraining from employing augmentation during the fine-tuning phase yields better results. Our experimental results confirm that the proposed methodology enhances retrieval and tagging performance metrics in two distinct scenarios: one where human-annotated tags are consistently available for all music tracks, and another where such tags are accessible only for a subset of tracks.
\end{abstract}

\section{Introduction}\label{sec:intro}
As web search engines have revolutionized the way individuals acquire information, advancements in music search systems hold the potential to become a pivotal force in the tailored delivery of music tracks to listeners and creators, thereby fostering the development of personalized music. With the proliferation of video content on social media platforms such as YouTube and TikTok, as well as in events like wedding celebrations, background music has emerged as a critical component. Consequently, there is an escalating demand for the ability to search for music that complements video content or aligns with specific occasions.
The growing popularity of music retrieval technology is evident in the widespread use of commercial applications such as Shazam (acquired by Apple), SoundHound, and Sony TrackID. These applications empower users to identify songs from brief samples captured by the device's microphone. However, akin to how web search engines serve purposes beyond mere webpage identification, music search should extend beyond song identification and facilitate the discovery of music that resonates with video content, advertisements for creative purposes, or personal preferences and moods for listening experiences.
Two fundamental capabilities of such music search systems encompass tag-based search through auto-tagging and similar music search via similarity-based music retrieval.

To deliver exceptional user experiences, search systems must exhibit remarkable accuracy, ensuring their outputs embody a deep comprehension of music and human music perception or recognition. The fundamental approach involves utilizing human-annotated tags for music. In auto-tagging, tags are classified based on a given music track using classification learning, and embeddings of music tracks with identical tags are learned to be similar through metric learning. However, due to the limitations and non-scalability of human supervision signals, it is imperative for the model to incorporate alternative signals to achieve enhanced performance. We address this challenge by learning from self-supervised signals, which derive from the music audio data itself.

Traditional music search technologies for similarity-based music retrieval rely on supervised learning, where learning signals originate from human-annotated tags \cite{MetricLearningISMIR2020}. Conversely, self-supervised learning has been employed for auto-tagging \cite{CLMR}. In this study, we present a model that integrates metric learning and self-supervised learning. We demonstrate that self-supervised learning is advantageous not only for auto-tagging but also for the similarity-based retrieval task. Furthermore, we introduce refined techniques to improve conventional self-supervised learning methods.

What is an intuitive explanation for our self-supervised signals? The similarity between music tracks is typically defined by their global similarity, which considers how closely related their global attributes are \cite{MetricLearningISMIR2020}. Auto-tagging performance is assessed based on the ability to infer global tags from each music track \cite{MetricLearningISMIR2020,MusicTagTrans,EvalMusicTag}. Our neural network aims to extract such global attribute features without relying solely on manually annotated tags. We formulate learning signals under the assumption that excerpts from the same track are more likely to possess similar global attribute features compared to excerpts from different tracks. Additionally, we assume that the global attribute features of a track remain relatively unchanged even after applying augmentation transformations, such as band-pass filtering or pitch shifting. Since the learning signal is based on supervision that does not necessitate human annotation but rather relies on annotations derived from the audio data itself, this learning approach is referred to as self-supervised learning.

To take advantage of self-supervision signals, we need careful consideration of the design of layers for self-supervision and metric learning. 
Given that global attribute features are more directly relevant to metric learning embeddings than classification probabilities, we meticulously determine metric learning embeddings comes right after the layer whose output feature is learned by self-supervised signals. We also carefully consider where to apply normalization operations and put that operations after branching to the supervised loss function head to avoid affecting the self-supervised loss function head.

Our self-supervised loss diverges from conventional self-supervised losses in several aspects. Self-supervised learning is frequently introduced in the context of representation learning, wherein the acquired representation, or feature, is fixed (the learned neural network is frozen), and the representation is employed for other tasks during the so-called fine-tuning phase  \cite{SimCLR,CLMR}. In this paper, we utilize self-supervised learning to enhance task performance and propose adapted learning techniques. Specifically, 1) during the fine-tuning phase, the neural network is not frozen, allowing the entire network to be trained to capitalize on its expressivity. 2) Self-supervised learning signals are employed even in the fine-tuning phase. 3) Augmentation is omitted for self-supervised learning during the fine-tuning phase, enabling our neural network to be trained with higher quality data. Overall, we consider the self-supervised signal as an auxiliary loss in relation to the primary metric learning loss, which improves performance compared to employing the standard self-supervised approach.

To further leverage the self-supervised signals, we empirically demonstrate that our method is also effective in addressing semi-supervised scenarios where obtaining human-annotated tags for music tracks is expensive and tags may not always be available for all tracks used in training models.

Our primary contributions can be summarized as follows:
\begin{itemize}
\item We propose a model that employs self-supervised learning to boost the performance of music similarity-based retrieval in both supervised and semi-supervised contexts.
\item We introduce a self-supervised auxiliary loss for music similarity-based retrieval and music auto-tagging, which serves to augment the outcomes in comparison to the conventional self-supervised approach within the supervised scenario.
\end{itemize}

\section{Methodology}\label{sec:methodology}
\subsection{Problem Setting}
Let us consider a dataset 
$$\mathcal{D}=\left\{(\boldsymbol{x}_{k}, \boldsymbol{y}_{k})\right\}_{k=1}^{N_{{\rm label}}} \bigcup \left\{\boldsymbol{x}_{k}\right\}_{k=1}^{N_{{\rm unlabel}}},$$
a set of $N_{{\rm label}}$ pairs of a music track $\boldsymbol{x}_{k} \in \mathcal{X}$ and its multi-tag $\boldsymbol{y}_{k} \in \mathcal{Y}$ and a set of  $N_{{\rm unlabel}}$ music tracks $\boldsymbol{x}_{k} \in \mathcal{X}$.
Our goal is to learn a mapping $F_{{\rm sim}}\colon \mathcal{X} \to \mathcal{Z}$ given $\mathcal{D}$, where $F_{{\rm sim}}(\boldsymbol{x}_{k})\in \mathcal{Z} \subset \mathbb{R}^D $ is an embedding vector, and some distance in the latent space $\mathcal{Z}$ captures similarity of data points $\boldsymbol{x}_{k}\in \mathcal{X}$. This is for the similarity-based retrieval task.
Our goal is also to learn a mapping $F_{{\rm tag}}\colon \mathcal{X} \to \mathcal{Y}$ given $\mathcal{D}$, where $F_{{\rm tag}}(\boldsymbol{x}_{k})\in \mathcal{Y} \subset [0,1]^T$ is a probability vector whose $t$-th element is the probability that $t$-th tag is assigned to $\boldsymbol{x}_{k}$. This is for the auto-tagging task.

\subsection{Outline}
\begin{figure*}[t!]
\centering
   \includegraphics[width=0.7\textwidth, trim={0cm 0cm 0cm 0cm}, clip]{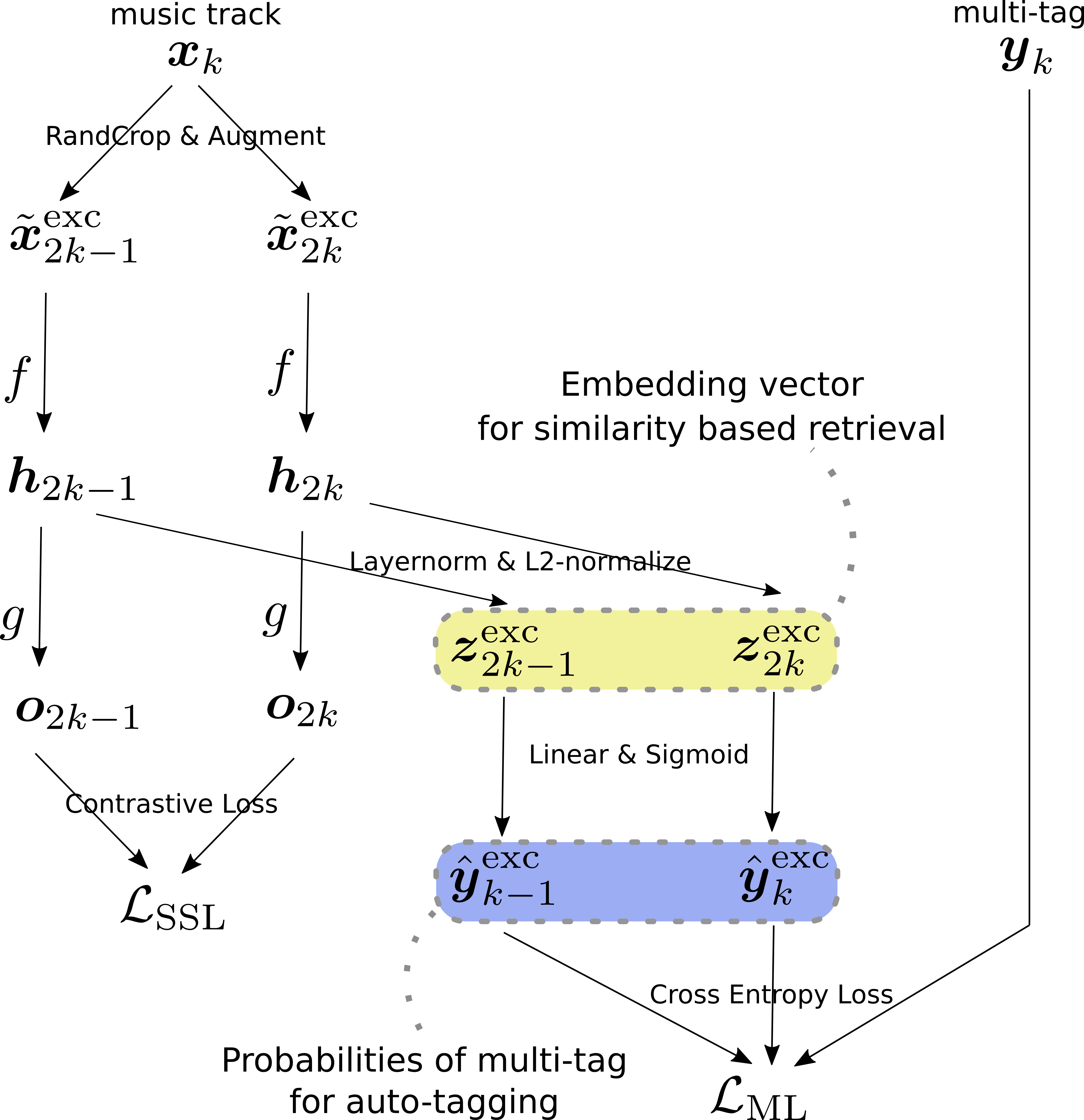}
  \caption{{\bf  Model overview.} For each batch comprising pairs of a music track $\boldsymbol{x}$ and its corresponding multi-tag $\boldsymbol{y}$, the music tracks undergo transformations (indicated by arrows) to compute the self-supervised learning loss $\mathcal{L}_{{\rm SSL}}$ and the metric learning loss $\mathcal{L}_{{\rm ML}}$. The losses are used to define the overall loss function $\mathcal{L}_{{\rm SSML}}=
\lambda\mathcal{L}_{{\rm SSL}}+\mathcal{L}_{{\rm ML}}$ (Eq~(\ref{overall_loss})) to train our proposed model. After training the model, given a music track $\boldsymbol{x}$, the embedding vector $\boldsymbol{z}^{\rm exc}$ and the estimated probabilities of multi-tag $\hat{\boldsymbol{y}}^{\rm exc}$ are used for similarity-based retrieval and auto-tagging, respectively.}
  \label{fig:overview} 
\end{figure*}

Figure.\ref{fig:overview} shows our model's overview. 
Instead of learning $F_{{\rm sim}}$ or $F_{{\rm tag}}$ directly, our model learns mappings whose inputs are excerpts $\boldsymbol{x}^{{\rm exc}} \in \mathcal{X}^{{\rm exc}}$,  cropped from music tracks, following previous work \cite{MetricLearningISMIR2020}. Formally, our model learns a mapping $f_{{\rm sim}} \colon \mathcal{X}^{{\rm exc}} \to \mathcal{Z}$, and we define 
\begin{equation}
F_{{\rm sim}}(\boldsymbol{x}_{k})=\operatorname{Aggregate}_{{\rm sim}}\left(\left(f_{{\rm sim}}(\boldsymbol{x}^{{\rm exc}}_{k,e})\right)_{e=1}^{E}\right),
\end{equation}
where $\left(\boldsymbol{x}^{{\rm exc}}_{k,e}\right)_{e=1}^{E}$ is a sequence of excerpts cropped from a track $\boldsymbol{x}_{k}$, and $\operatorname{Aggregate}_{{\rm sim}}(\cdot)$ is the arithmetic mean operation followed by division by $\ell^2$-norm.  Similarly, our model also learns a mapping $f_{{\rm tag}} \colon \mathcal{X}^{{\rm exc}} \to \mathcal{Y}$, and we define 
\begin{equation}
F_{{\rm tag}}(\boldsymbol{x}_{k})=\operatorname{Aggregate}_{{\rm tag}}\left(\left(f_{{\rm tag}}(\boldsymbol{x}^{{\rm exc}}_{k,e})\right)_{e=1}^{E}\right),
\end{equation}
where $\operatorname{Aggregate}_{{\rm tag}}(\cdot)$ is the arithmetic mean operation followed by the softmax operation.
 In experiments, excerpts are non-overlapping sliding windows in each track.
Similarity learning (metric learning) is achieved by tagging (classification) based methodology, as revealed in prior studies \cite{ClassificationMetricLearning,SoftTripleLoss}. Thus, our model learns $f_{{\rm tag}}$ such that 
\begin{equation}
f_{{\rm tag}}(\boldsymbol{x}^{{\rm exc}}_{k,e})=\sigma\left(Wf_{{\rm sim}}(\boldsymbol{x}^{{\rm exc}}_{k,e})\right),
\end{equation}
where $W \in \mathbb{R}^{T\times D}$ and $\sigma$ denotes the sigmoid activation.  Model architectures for similarity-based retrieval and auto-tagging are mostly shared in this formulation, so it is advantageous in practice in terms of time, memory, and storage in training and inference phases, particularly when using functionalities of both similarity-based retrieval and auto-tagging. In Sections \ref{subsec:self-supervised}, \ref{subsec:self-supervised metric}, we explain how to train $f_{{\rm sim}}$ and $W$ (thus $f_{{\rm tag}}$) in detail, where $f_{{\rm sim}}$ is defined as 
\begin{equation}
f_{{\rm sim}}(\cdot)=\frac{{\rm LN}(f(\cdot))}{\left\|{\rm LN}(f(\cdot))\right\|_2},
\end{equation}
where {\rm LN} denotes layer normalization \cite{LayerNorm}.
Then our goal in the Sections \ref{subsec:self-supervised} and \ref{subsec:self-supervised metric} boils down to learning $f$ and $W$, where we choose to use the SampleCNN architecture for $f$ \cite{lee2018samplecnn}. $f$ is trained using a self-supervised learning loss and a metric learning loss, whereas $W$ is trained only using a metric learning loss. Since inner product is the distance metric between each row of $W$ and $f_{{\rm sim}}(\boldsymbol{x}^{{\rm exc}}_{k,e})$, we use inner product as the distance metric in the similarity space when conducting similarity-based retrieval.

\subsection{Self-supervised Learning}\label{subsec:self-supervised}
Consider a mini-batch $\left\{\boldsymbol{x}_{k}\right\}_{k=1}^{B}$ from the dataset $\mathcal{D}$ and a set of augmentation operations $\mathcal{A}$ (See Section~\ref{subsec:model_config} for the choice of $\mathcal{A}$ in experiments). We follow the Contrastive Learning of Musical Representation (CLMR) \cite{CLMR}, which uses the SimCLR framework for self-supervised learning \cite{SimCLR}. 
For each mini-batch training, we sample two augmentation operations $a, a^{\prime} \sim \mathcal{A}$ and compute the following transformations. 

\begin{align} 
\tilde{\boldsymbol{x}}^{{\rm exc}}_{2 k-1} &=a\left(\operatorname{RandCrop}\left(\boldsymbol{x}_{k}\right)\right) \\ \boldsymbol{h}_{2 k-1} &=f\left(\tilde{\boldsymbol{x}}^{{\rm exc}}_{2 k-1}\right) \\ \boldsymbol{o}_{2 k-1} &=g\left(\boldsymbol{h}_{2 k-1}\right) \\
\tilde{\boldsymbol{x}}^{{\rm exc}}_{2 k} &=a^{\prime}\left(\operatorname{RandCrop}\left(\boldsymbol{x}_{k}\right)\right) \\ \boldsymbol{h}_{2 k} &=f\left(\tilde{\boldsymbol{x}}^{{\rm exc}}_{2 k}\right) \\ \boldsymbol{o}_{2 k} &=g\left(\boldsymbol{h}_{2 k}\right) 
\end{align}
where a pair $(\tilde{\boldsymbol{x}}^{{\rm exc}}_{2 k-1}, \tilde{\boldsymbol{x}}^{{\rm exc}}_{2 k})$ is referred to as a {\it positive pair}, and random crop (denoted as $\operatorname{RandCrop}(\cdot)$) and augmentation operations are assumed to preserve the global attributes. The random crop refers to cropping an excerpt from a music track, where the excerpt position in a music track is drawn uniformly from all possible positions. For the architecture of $g$, we use a linear layer followed by a ReLU layer followed by a linear layer, where no bias term is used in the linear layers.

Given a set $\left\{\tilde{\boldsymbol{x}}^{{\rm exc}}_{l}\right\}_{l=1}^{2B}$ including a positive pair of examples $\tilde{\boldsymbol{x}}^{{\rm exc}}_{i}$ and $\tilde{\boldsymbol{x}}^{{\rm exc}}_{j}$, the contrastive prediction task aims to identify $\tilde{\boldsymbol{x}}^{{\rm exc}}_{j}$ in $\left\{\tilde{\boldsymbol{x}}^{{\rm exc}}_{l}\right\}_{l \neq i}$ for a given $\tilde{\boldsymbol{x}}^{{\rm exc}}_{i}$.
Formally, letting $\operatorname{sim}(\boldsymbol{u}, \boldsymbol{v})=\boldsymbol{u}^{\top} \boldsymbol{v} /\|\boldsymbol{u}\|_2\|\boldsymbol{v}\|_2$, a contrastive loss function can be defined for a contrastive prediction task as
\begin{equation}
L_{{\rm SSL}}(i, j)=-\log \frac{\exp \left(\operatorname{sim}\left(\boldsymbol{o}_{i}, \boldsymbol{o}_{j}\right) / \tau\right)}{\sum_{l=1}^{2 B} \mathbbm{1}_{[l \neq i]} \exp \left(\operatorname{sim}\left(\boldsymbol{o}_{i}, \boldsymbol{o}_{l}\right) / \tau\right)},
\end{equation}
where $\tau$ is a temperature parameter set to the default value proposed in SimCLR \cite{SimCLR}.
$L_{{\rm SSL}}(i, j)$ is computed for all augmented pairs, i.e., $(i, j)\in \{(2k-1,2k)\}_{k=1}^{B} \bigcup \{(2k,2k-1)\}_{k=1}^{B}$ and averaged, yielding the overall loss function
\begin{equation}
\mathcal{L}_{{\rm SSL}}=\frac{1}{2 B} \sum_{k=1}^{B}[L_{{\rm SSL}}(2 k-1,2 k)+L_{{\rm SSL}}(2 k, 2 k-1)].\label{SSL_loss}
\end{equation}

\subsection{Metric Learning with Self-supervised Auxiliary Loss}\label{subsec:self-supervised metric}
We propose to combine classification-based metric learning with self-supervised learning.
Layer normalization (denoted by ${\rm LN}(\cdot)$) is applied to $\boldsymbol{h}_{i}$, followed by normalization with $\ell^2$-norm to yield an embedding vector $\boldsymbol{z}^{\rm exc}_{i}\in \mathbb{R}^D$ for similarity-based retrieval. Formally,  \begin{equation}
\boldsymbol{z}^{\rm exc}_{i}=\frac{{\rm LN}(\boldsymbol{h}_{i})}{\left\|{\rm LN}(\boldsymbol{h}_{i})\right\|_2}.
\end{equation}
$\boldsymbol{z}^{\rm exc}_{i}$ is then multiplied by $W$, followed by element-wise sigmoid activation to produce classification output $\hat{\boldsymbol{y}}^{\rm exc}_{i}$, i.e.,
\begin{equation}
\hat{\boldsymbol{y}}^{\rm exc}_{i}=\sigma\left(W\boldsymbol{z}^{\rm exc}_{i}\right).
\end{equation}
We use binary cross entropy loss for each tag and average them:
\begin{equation}
L_{{\rm ML}}(i)=\frac{1}{T}\sum_{t}^{T}\left[-\boldsymbol{y}_{i}[t] \log \left(\hat{\boldsymbol{y}}^{\rm exc}_{i}[t]\right)-\left(1-\boldsymbol{y}_{i}[t]\right) \log \left(1-\hat{\boldsymbol{y}}^{\rm exc}_{i}[t]\right)\right].
\end{equation}
Let $\mathcal{K}_{\rm label}$ be an index set such that $\left\{\boldsymbol{x}_{k}\colon k \in \mathcal{K}_{\rm label}\subseteq \{1,2, ..., B\} \right\}$ is the set of all the labeled samples in $\left\{\boldsymbol{x}_{k}\right\}_{k=1}^{B}$. $L_{{\rm ML}}(i)$ is computed for the samples in the labeled subset and averaged, yielding the loss function
\begin{equation}
\mathcal{L}_{{\rm ML}}=\frac{1}{\lvert\mathcal{K}_{\rm label}\rvert} \sum_{k \in \mathcal{K}_{\rm label}}[L_{{\rm ML}}(2k-1)+L_{{\rm ML}}(2k)].
\end{equation}
Finally, the loss function for our proposed model is 
\begin{equation}
\mathcal{L}_{{\rm SSML}}=
\lambda\mathcal{L}_{{\rm SSL}}+\mathcal{L}_{{\rm ML}}.\label{overall_loss}
\end{equation}
Here $\lambda \in \mathbb{R}$ is a balancing factor between two losses $\mathcal{L}_{{\rm SSL}}$ and $\mathcal{L}_{{\rm ML}}$.

In practice, the self-supervised learning needs a longer training time, so we first train our model with  $\mathcal{L}_{{\rm SSL}}$ only, whose phase is referred to as pre-training phase. We then train with $\mathcal{L}_{{\rm SSML}}$, whose phase is referred to as fine-tuning phase.

\section{Experimental Setup}\label{sec:experiments}
\subsection{Dataset}
\subsubsection{MagnaTagATune dataset}
The MagnaTagATune dataset consists of 25,000 music
tracks from 6,622 unique songs \cite{MagnaTagATune}. We use top 50 tags and the same train-test split as in previous work \cite{CLMR}. We obtained the MagnaTagATune dataset using the code \url{https://github.com/Spijkervet/CLMR/blob/master/clmr/datasets/magnatagatune.py}, where the dataset itself is downloaded from \url{https://github.com/minzwon/sota-music-tagging-models/tree/master/split/mtat}.
\subsubsection{MTG-Jamendo dataset}
MTG-Jamendo contains 55,000 full audio tracks (320kbps, MP3) with 195 tags covering genre, instrument, and mood/theme \cite{bogdanov2019mtg}. The dataset comes with a pre-defined split based on the target tasks. We use the pre-defined split and the top 50 tags for training and evaluation. We obtained the MTG-Jamendo dataset from \url{https://github.com/MTG/mtg-jamendo-dataset}.

\subsection{Model Configurations}\label{subsec:model_config}
The set of augmentation operations $\mathcal{A}$ follows CLMR \cite{CLMR} for fair comparison. Specifically, the following operations are applied sequentially with probability $p$ to create an element of $\mathcal{A}$.
\begin{itemize}
\item polarity inversion ($p=0.8$)
\item additive Gaussian noise with decibel sampled uniformly from $[80, 40]$ ($p=0.01$)
\item gain with decibel sampled uniformly from $[-6, 0]$ ($p=0.3$)
\item low pass filtering or high pass filtering  chosen with the same probability, where their cut-off frequency is sampled uniformly from $[2200, 4000]$ Hz and $[200, 1200]$ Hz, respectively ($p=0.8$)
\item delayed signal added to the original signal with a volume factor of $0.5$ in which the delay time is randomly sampled from $\{200, 250, 300, ..., 500\}$ ms ($p=0.3$)
\item pitch shifting with shifting semitones sampled uniformly from $[-7, 7]$ ($p=0.6$)
\item reverb with the impulse response’s room size, reverberation, and damping
factor sampled uniformly from $[0, 100]$ ($p=0.6$)
\end{itemize}

We set the excerpt length to 59049 and audio sampling rate to 22.05 kHz following CLMR \cite{CLMR} for fair comparison.

To determine the value of $\lambda$ in Eq~(\ref{overall_loss}), we first introduce the base balancing factor $r$ of the two terms $\mathcal{L}_{\rm ML}$ and $\mathcal{L}_{\rm SSL}$. $r$ is defined to be $r=\mathcal{L}^{\rm only}_{\rm ML}/\mathcal{L}^{\rm only}_{\rm SSL}$, where $\mathcal{L}^{\rm only}_{\rm ML}$ and $\mathcal{L}^{\rm only}_{\rm SSL}$ are the converged loss values when the model is trained using either $\mathcal{L}_{\rm ML}$ or $\mathcal{L}_{\rm SSL}$, respectively, and all available labels are used when trained with $\mathcal{L}_{\rm ML}$. The values of $r$ were $22.00$ for MagnaTagATune dataset and $18.95$ for MTG-Jamendo dataset.
Then, the candidates for $\lambda$ in Eq~ (\ref{overall_loss}) were set to $\{\alpha / r : \alpha \in \{0.05, 0.1, 1, 10\}\}$. For conciseness, $\{\alpha / r : \alpha \in \{0.1, 1, 10\}\}$ for the MagnaTagATune dataset and $\{\alpha / r : \alpha \in \{0.05, 0.1, 1\}\}$ for the MTG-Jamendo dataset are shown in Tables \ref{table:mtat} and \ref{table:mtg}, respectively.

In our model's pre-training where only $\mathcal{L}_{\rm SSL}$ is used, the batch size is set to $48$, we employ the Adam optimizer with a learning rate of $0.0003$ and $\beta_1,\beta_2=(0.9, 0.999)$. The model is trained for $10,000$ and $1,000$ epochs for MagnaTagATune and MTG-Jamendo, respectively.

For our model's fine-training where the overall loss $\mathcal{L}_{\rm SSML}$ is used, the batch size is set to $48$. We use the Adam optimizer with a learning rate of $0.001$ and $\beta_1,\beta_2=(0.9, 0.999)$, in which the learning rate is multiplied by $0.1$ when the validation loss does not improve for $5$ epochs. We use a weight decay with a weight of $1.0e-6$, and the model is trained for $200$ epochs maximum. The training is stopped when the validation loss does not improve for $10$ epochs, which is referred to as early stopping.

\subsection{Evaluation Metrics}
\subsubsection{Similarity-based Retrieval}
To evaluate the similarity-based retrieval, we use the
recall@K (R@K) metric to measure retrieval quality following the standard evaluation setting in image retrieval \cite{ClassificationMetricLearning,SoftTripleLoss} and a music similarity-based retrieval model \cite{MetricLearningISMIR2020}. This metric is useful for evaluating search methods because it measures the quality of the top K retrieved results, which are more important and more likely to be seen by users than lower ranked retrieved results.

\subsubsection{Auto-tagging}
Music auto-tagging has been extensively studied, and diverse model architectures has been developed \cite{MetricLearningISMIR2020,MusicTagTrans,EvalMusicTag}. 
We follow the standard benchmarking and evaluation criteria
and report average tag-wise area under the receiver operating characteristic curve (ROC-AUC) and average precision
(PR-AUC) scores to measure tag-based retrieval performance.
\subsection{Baseline Methods}
We compare our model with what we call the inception model, a state-of-the-art model for similarity-based retrieval and auto-tagging \cite{MetricLearningISMIR2020}. We also compare our model with CLMR \cite{CLMR}, a model for auto-tagging which uses SimCLR as self-supervised learning for pre-training \cite{SimCLR}.

\subsection{Variations of Learning Techniques}\label{subsec:LearningTechniques}
In this section, we discuss three learning techniques that define the variations of our proposed methods and the baseline approaches.

\subsubsection{Fine-tune Augment}
Fine-tune augment involves applying augmentation operations (as detailed in Section~\ref{subsec:model_config}) during the fine-tuning phase. Note that the inception model and CLMR do not utilize this technique.

\subsubsection{Fine-tune Contrastive}
Fine-tune contrastive entails conducting contrastive self-supervised learning, where the loss is given by Eq~$(\ref{SSL_loss})$, during the fine-tuning phase. It is noteworthy that neither the inception model nor CLMR employ this technique.

\subsubsection{Load Pre-train}
Load pre-train refers to loading the pre-trained model's weights during the fine-tuning phase. The pre-training is executed using the contrastive self-supervised loss specified by Eq~$(\ref{SSL_loss})$. It is pertinent to mention that while CLMR uses this technique, the inception model does not. Moreover, in our proposed methods, we do not freeze the models, even when the pre-trained weights are loaded.

\section{Results}
\subsection{Supervised: Scenario where tags are always available for music tracks}\label{result_supervised}
 Table~\ref{table:mtat} shows the results for the supervised scenario of the MagnaTagATune dataset, where ``Fine-tune Augment'', ``Fine-tune Contrastive'', and ``Load Pre-train'' are learning techniques that characterize the variations of especially our proposed methods (See Section~\ref{subsec:LearningTechniques}). Ours G  outperformed the previous methods, inception and CLMR, on both similarity-based retrieval and auto-tagging tasks. Ours A uses the same learning algorithm as that of inception except for the input representation and network architectures, the results of which suggest that the changes do not always lead to higher performance. Ours B is ``fine-tune augment'' added to ours A, which slightly improved some metrics and slightly degraded some other metrics, although augmentation is usually an effective strategy. Ours C, ``Load Pre-train'' added to ours A, improves the performance decently. ``Load Pre-train'' is the same strategy as CLMR, but ours C outperforms it presumably because ours does not freeze the pre-trained network and takes advantage of the expressivity of the pre-trained network.  We found that conducting self-supervised learning while fine-tuning boosts the performance (ours F, G), especially when no augmentation is performed while fine-tuning (ours G).

\begin{table*}[]
  \caption{{\bf Results for supervised scenario of MagnaTagATune dataset.} Ours A-I are compared with baseline methods inception and CLMR. ``Fine-tune Augment'', ``Fine-tune Contrastive'', and ``Load Pre-train'' are learning techniques that characterize the variations of especially our proposed methods (See Section~\ref{subsec:LearningTechniques}). Ours G generally achieves the highest scores for the both tasks.}
  \label{table:mtat}
   \centering
   \begin{tabular}{ccccccccccccc}
    \hline
    \multirow{2}{*} & \multirow{2}{*}{Models} & Fine-tune & Fine-tune & Load & \multicolumn{4}{c}{Similarity-based retrieval} & \multicolumn{2}{c}{Auto-tagging AUC}  \\
     & & Augment & Contrastive & Pre-train & R@1 & R@2 & R@4 & R@8 & ROC & PR \\
    \hline \hline
    \multirow{1}{*}
    & inception & & & &  51.7 & 66.3 & 78.3 & 87.5 & 0.905 & 0.375 \\
    \multirow{1}{*}
    & CLMR & & & \checkmark &  &  &  &  & 0.894 & 0.368 \\
    \hline
    \multirow{9}{*}
       & ours A & & & & 52.1 & 66.4 & 78.7 & 87.6 & 0.901 & 0.371 \\
    & ours B & \checkmark & & & 51.0 & 66.1 & 78.8 & 87.8 & 0.900 & 0.373 \\
    & ours C & & & \checkmark & 52.4 & 66.8 & 79.3 & {\bf 88.6} & 0.904 & 0.377 \\
    & ours D ($\alpha$=0.1) & \checkmark & \checkmark & \checkmark & 52.2 & 66.7 & 78.8 & 88.2 & 0.905 & 0.381 \\
    & ours E ($\alpha$=0.1) & & \checkmark & \checkmark & {\bf 53.0} & 67.1 & 78.8 & 88.1 & {\bf 0.906} & 0.381 \\
    & ours F ($\alpha$=1) & \checkmark & \checkmark & \checkmark & {\bf 53.0} & 66.7 & 79.2 & 88.3 & 0.905 & 0.381 \\
       & ours G ($\alpha$=1) & & \checkmark & \checkmark & {\bf 53.0} & {\bf 67.5} & {\bf 79.4} & 88.5 & {\bf 0.906} & {\bf 0.382} \\
    & ours H ($\alpha$=10) & \checkmark & \checkmark & \checkmark & 52.3 & 66.6 & 78.5 & 87.7 & 0.891 & 0.352 \\
    & ours I ($\alpha$=10) & & \checkmark & \checkmark & 52.8 & 66.6 & 78.6 & 87.7 & 0.897 & 0.361 \\
    \hline
   \end{tabular}
\end{table*}
  Table~\ref{table:mtg} shows the results for the supervised scenario of MTG-Jamendo dataset. Ours M was the most effective for similarity-based retrieval and had comparable performance to inception in terms of auto-tagging. Note that ours M and G use the same methodology (ours with ``Fine-tune Contrastive'' and ``Load Pre-train'') and this methodology is the most effective consistently across different datasets.
\begin{table*}[]
  \caption{{\bf Results for supervised scenario of MTG-Jamendo dataset.} Ours J-O are compared with baseline methods inception and CLMR. ``Fine-tune Augment'', ``Fine-tune Contrastive'', and ``Load Pre-train'' are learning techniques that characterize the variations of especially our proposed methods (See Section~\ref{subsec:LearningTechniques}). Ours M generally achieves the highest scores for the similarity-based retrieval task and inception achieves highest scores for the auto-tagging task. Note that ours M and G (in Table~\ref{table:mtat}) use the same methodology (ours with ``Fine-tune Contrastive'' and ``Load Pre-train'') and among ours this methodology is the most effective consistently across different datasets.}
  \label{table:mtg}
   \centering
   \begin{tabular}{ccccccccccccc}
    \hline
    \multirow{2}{*} & \multirow{2}{*}{Models} & Fine-tune & Fine-tune & Load & \multicolumn{4}{c}{Similarity-based retrieval} & \multicolumn{2}{c}{Auto-tagging AUC}  \\
     & & Augment & Contrastive & Pre-train & R@1 & R@2 & R@4 & R@8 & ROC & PR \\
    \hline \hline
    \multirow{1}{*}
    & inception & & & & 47.5 & 61.2 & 73.5 & 83.6 & {\bf 0.829} & {\bf 0.292} \\
    \hline
    \multirow{6}{*}
    & ours J $\alpha$=0.05 & \checkmark & \checkmark & \checkmark & 49.3 & 62.3 & 73.7 & 83.5 & 0.825 & 0.285 \\
    & ours K $\alpha$=0.05 &  & \checkmark & \checkmark & 52.1 & 64.5 & 75.7 & 84.6 & 0.826 & 0.286 \\
    & ours L $\alpha$=0.1 & \checkmark & \checkmark & \checkmark & 49.7 & 62.5 & 74.2 & 83.8 & 0.826 & 0.288 \\
    & ours M $\alpha$=0.1 &  & \checkmark & \checkmark & {\bf 52.3} & {\bf 65.1} & {\bf 76.0} & {\bf 84.8} & 0.828 & 0.287 \\
    & ours N $\alpha$=1 & \checkmark & \checkmark & \checkmark & 47.6 & 60.2 & 72.2 & 82.4 & 0.822 & 0.278 \\
    & ours O $\alpha$=1 &  & \checkmark & \checkmark & 50.0 & 62.2 & 73.5 & 82.8 & 0.825 & 0.285 \\
    \hline
   \end{tabular}
\end{table*}

\subsection{Semi-supervised:  Scenario where tags are not always available for music tracks}
\begin{figure*}[]
\centering
   \subfloat[R@1]{
    \includegraphics[width=0.32\textwidth, trim={0cm 0cm 0cm 0cm}, clip]{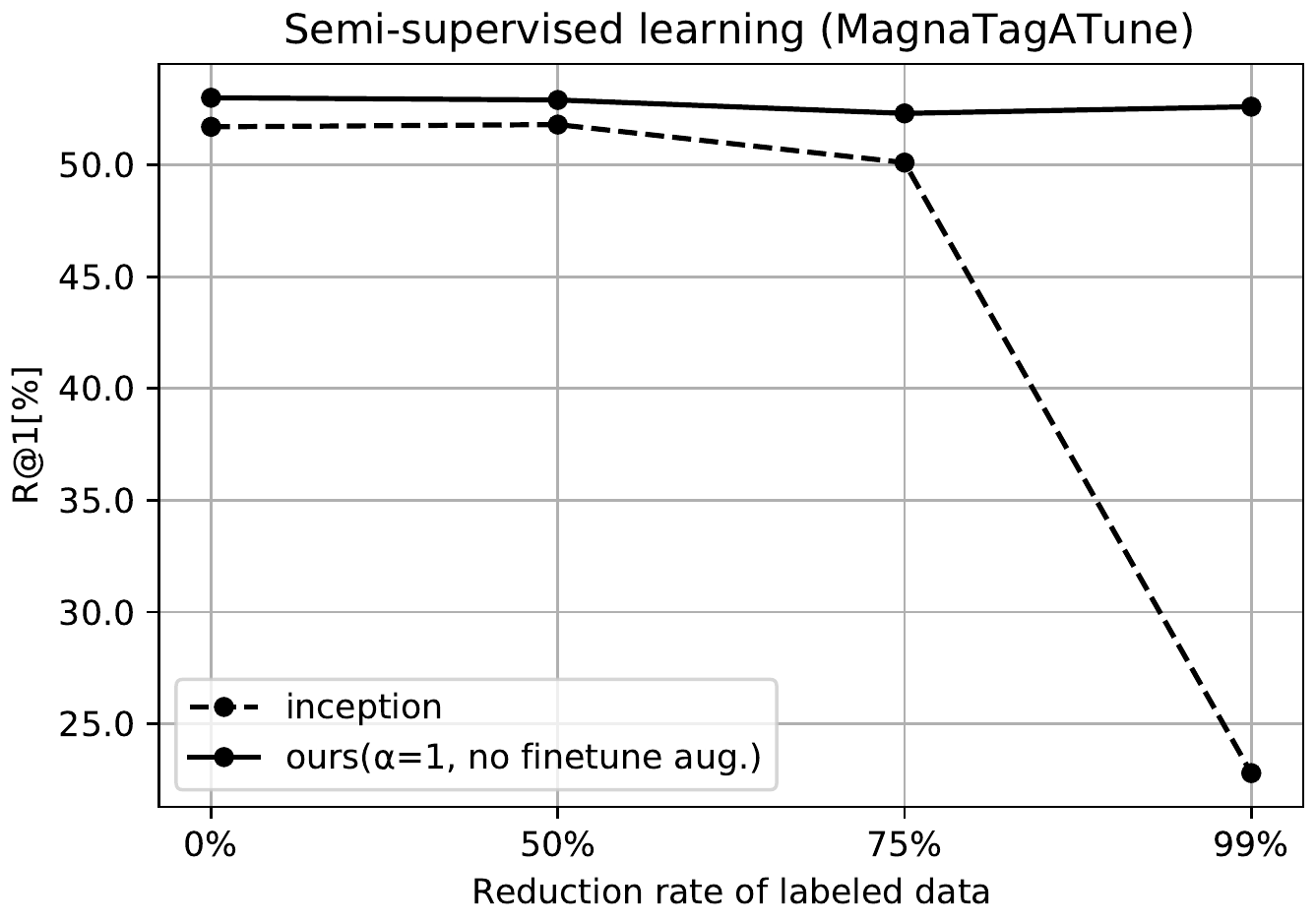}
  }
  \subfloat[R@2]{
  \includegraphics[width=0.32\textwidth, trim={0cm 0cm 0cm 0cm}, clip]{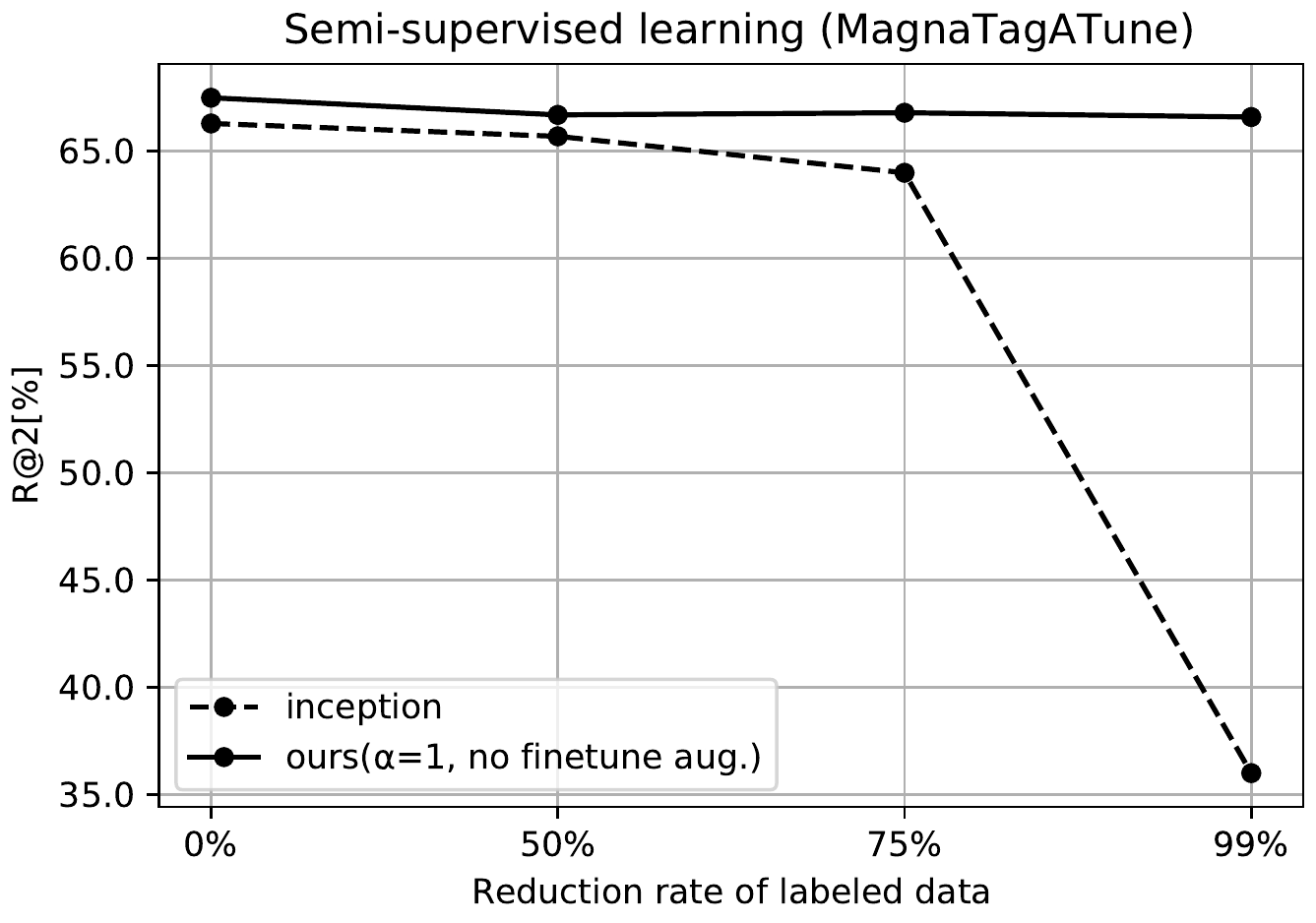}
  \label{fig:obj_eval}  }
 \subfloat[R@4]{
    \includegraphics[width=0.32\textwidth, trim={0cm 0cm 0cm 0cm}, clip]{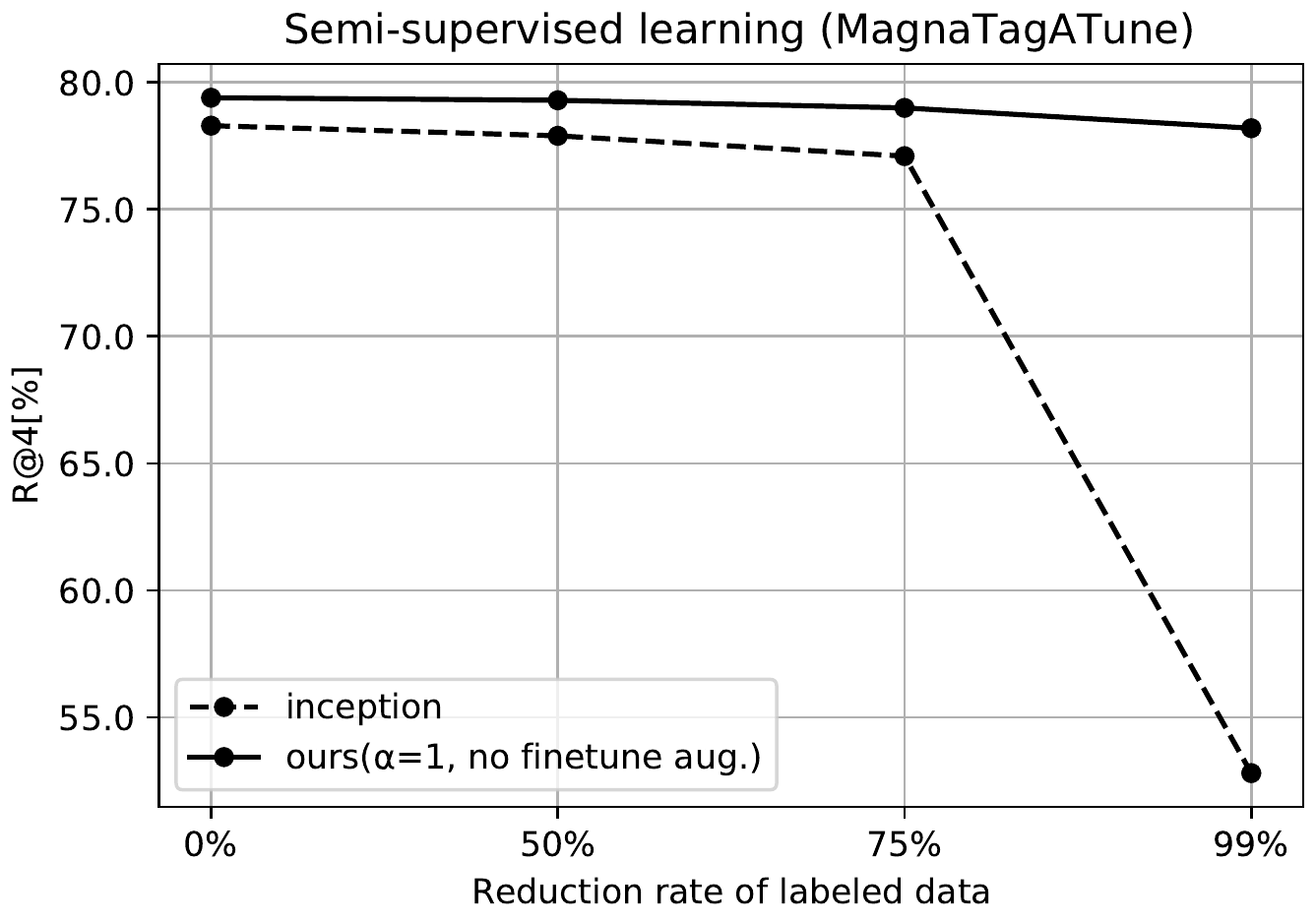}
  \label{fig:human_eval}
  }\\
     \subfloat[R@8]{
    \includegraphics[width=0.32\textwidth, trim={0cm 0cm 0cm 0cm}, clip]{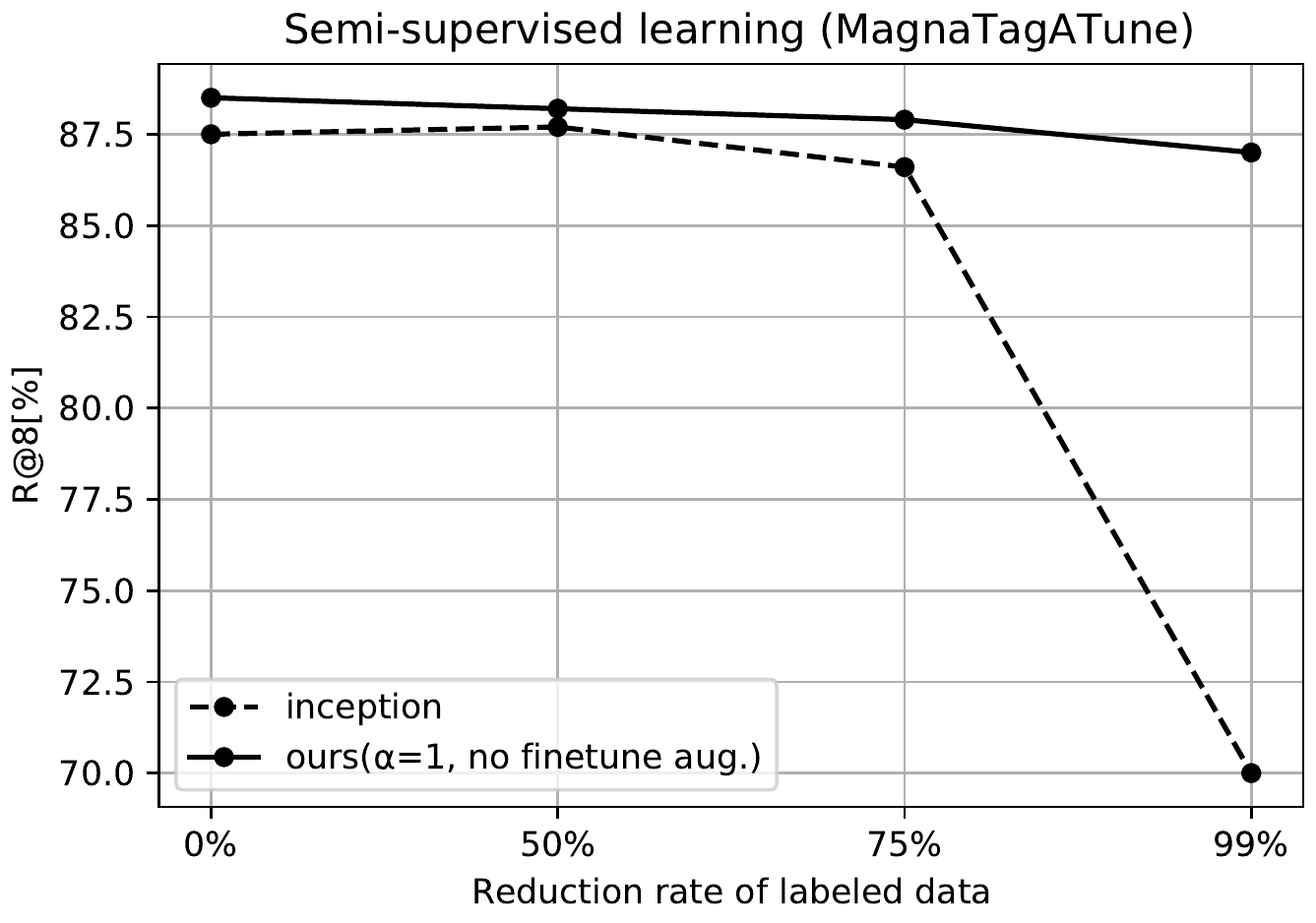}
  }
  \subfloat[ROC]{
  \includegraphics[width=0.32\textwidth, trim={0cm 0cm 0cm 0cm}, clip]{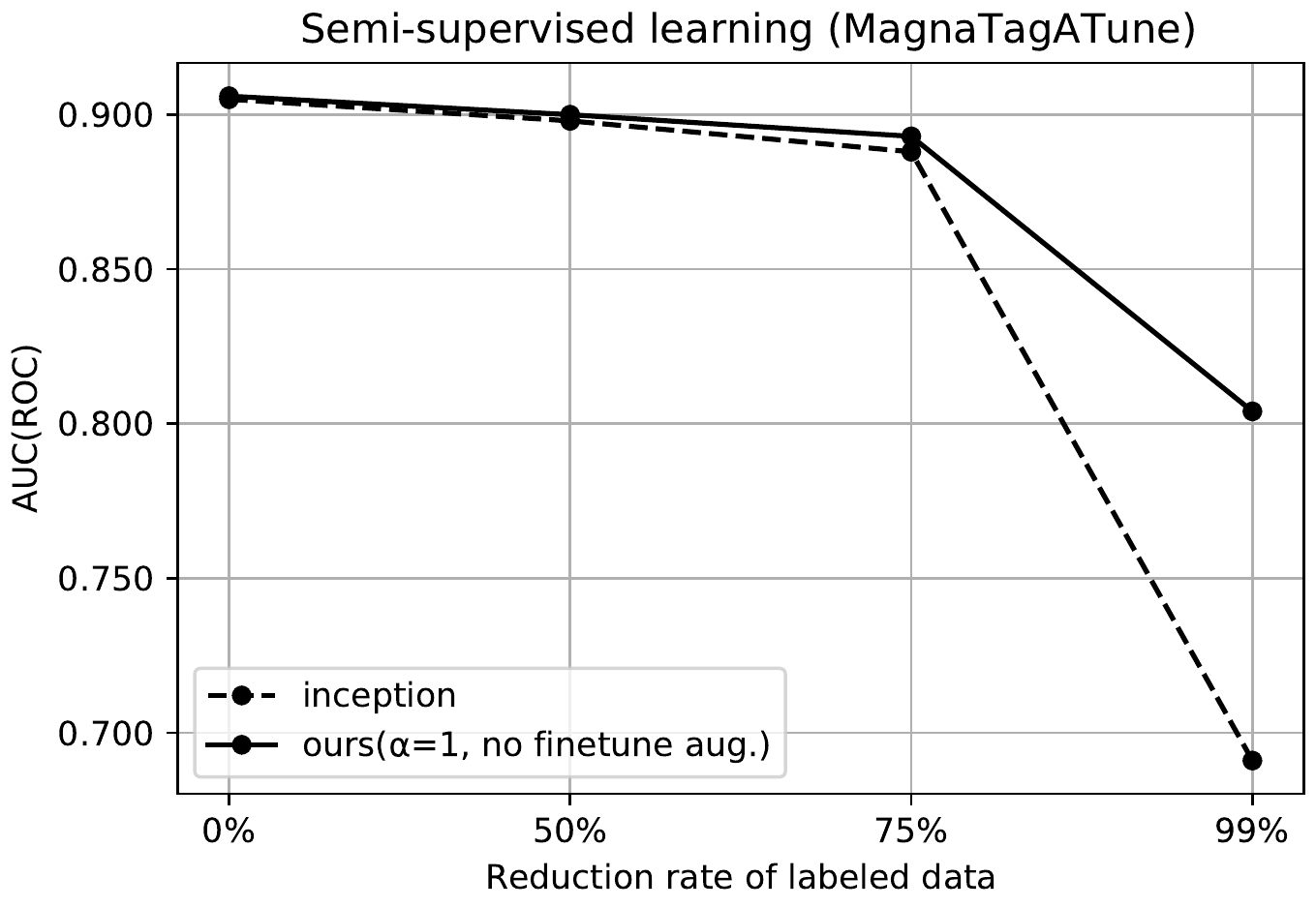}
  \label{fig:obj_eval}  }
 \subfloat[PR]{
    \includegraphics[width=0.32\textwidth, trim={0cm 0cm 0cm 0cm}, clip]{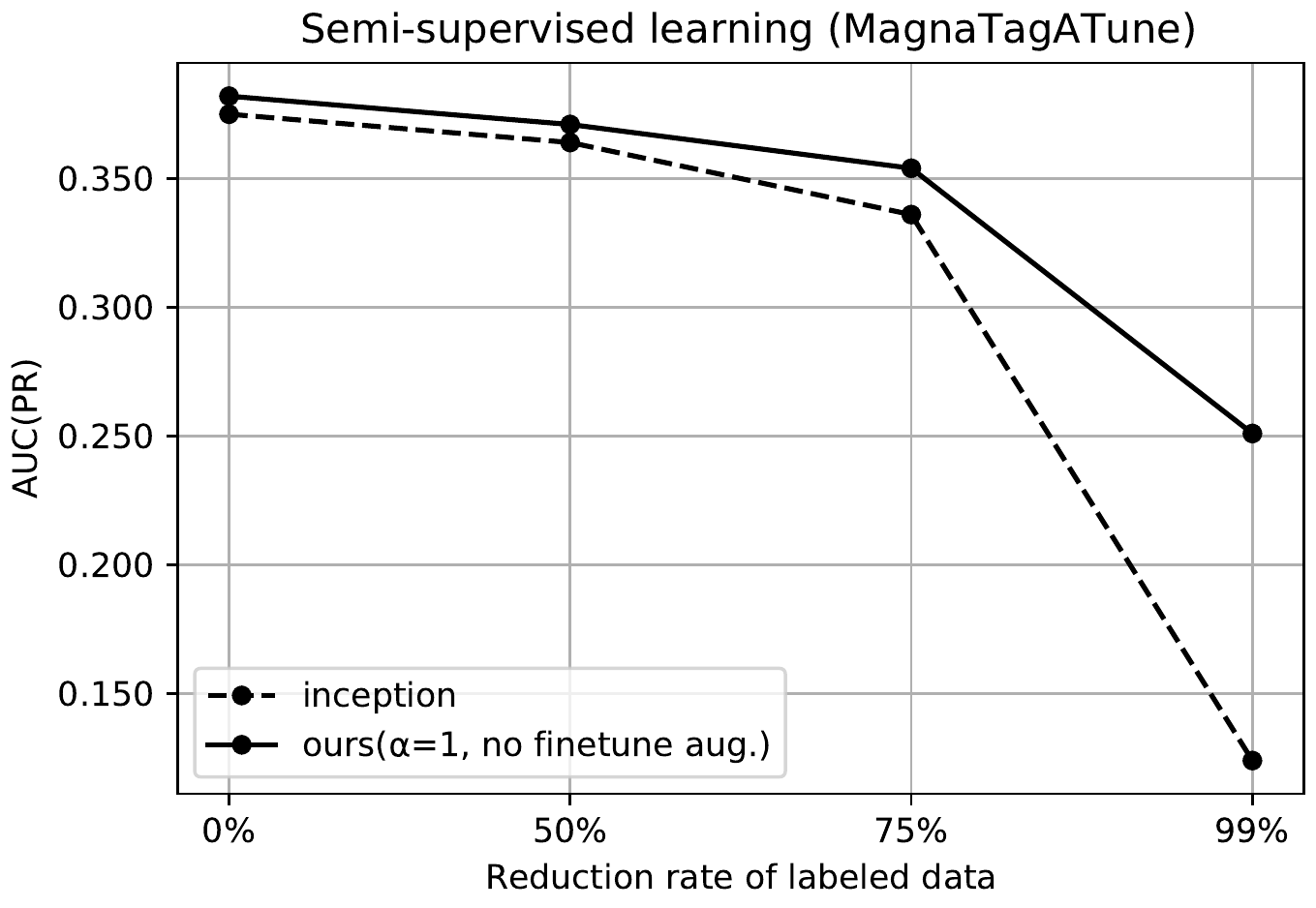}
  \label{fig:human_eval}
  }
  \caption{{\bf Results for semi-supervised scenario of MagnaTagATune dataset.} (a-d) Similarity-based retrieval R@K results. (e-f) Auto-tagging AUC results.}
     \label{fig:mtat}
\end{figure*}

We simulate the semi-supervised setting by reducing the rate of tags to be used.
 Fig~\ref{fig:mtat} shows the results for the semi-supervised scenario of the MagnaTagATune dataset. 
Compared to the inception model, the performance gain of our model becomes larger as the amount of labeled data decreases. For similarity-based retrieval (a-d), the performance of our model only degraded slightly even with a 99\% reduction in labeled data (i.e., with only 1\% of labeled data).

\begin{figure*}[]
\centering
   \subfloat[R@1]{
    \includegraphics[width=0.32\textwidth, trim={0cm 0cm 0cm 0cm}, clip]{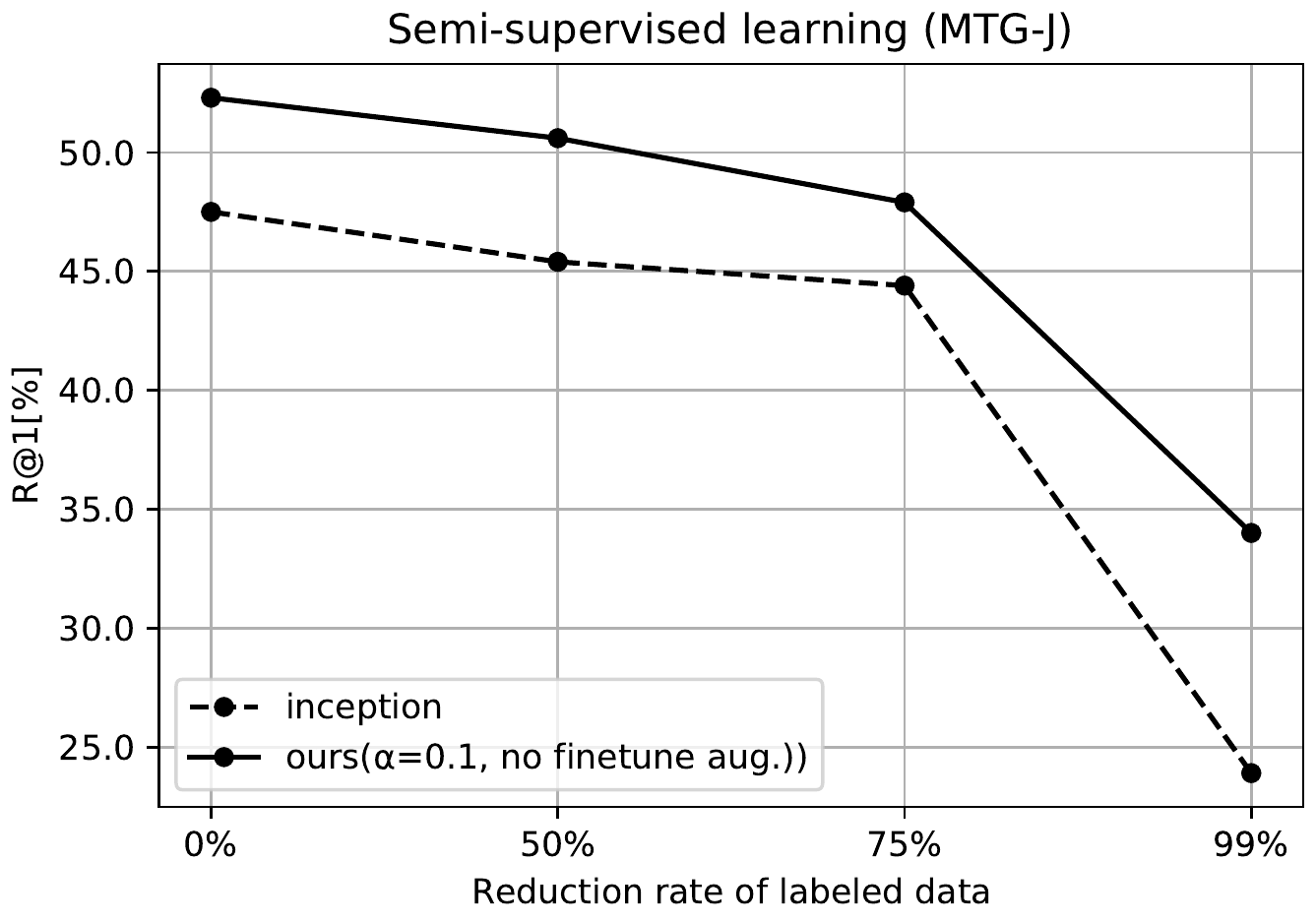}
  }
  \subfloat[R@2]{
  \includegraphics[width=0.32\textwidth, trim={0cm 0cm 0cm 0cm}, clip]{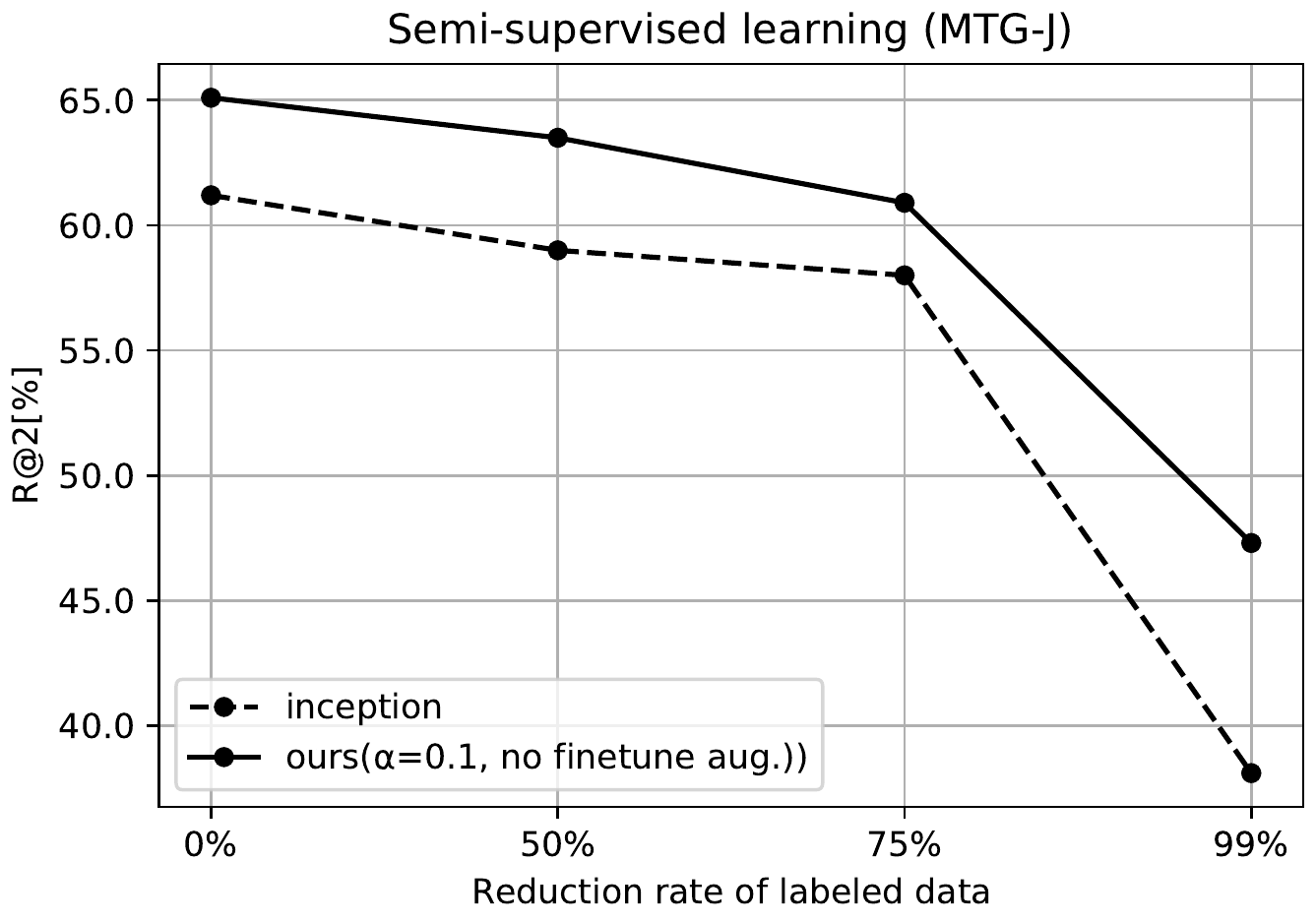}
  \label{fig:obj_eval}  }
 \subfloat[R@4]{
    \includegraphics[width=0.32\textwidth, trim={0cm 0cm 0cm 0cm}, clip]{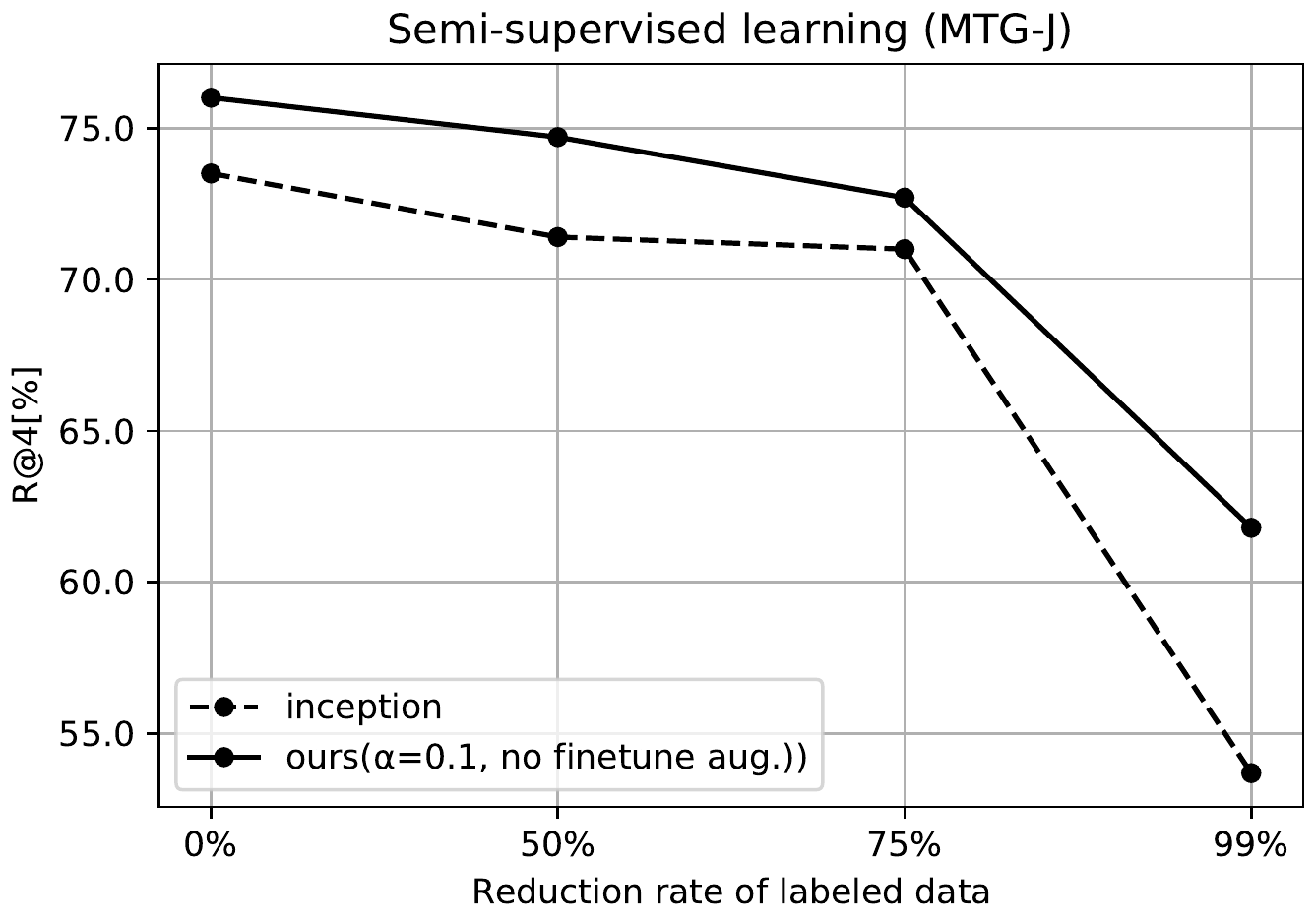}
  \label{fig:human_eval}
  }\\
     \subfloat[R@8]{
    \includegraphics[width=0.32\textwidth, trim={0cm 0cm 0cm 0cm}, clip]{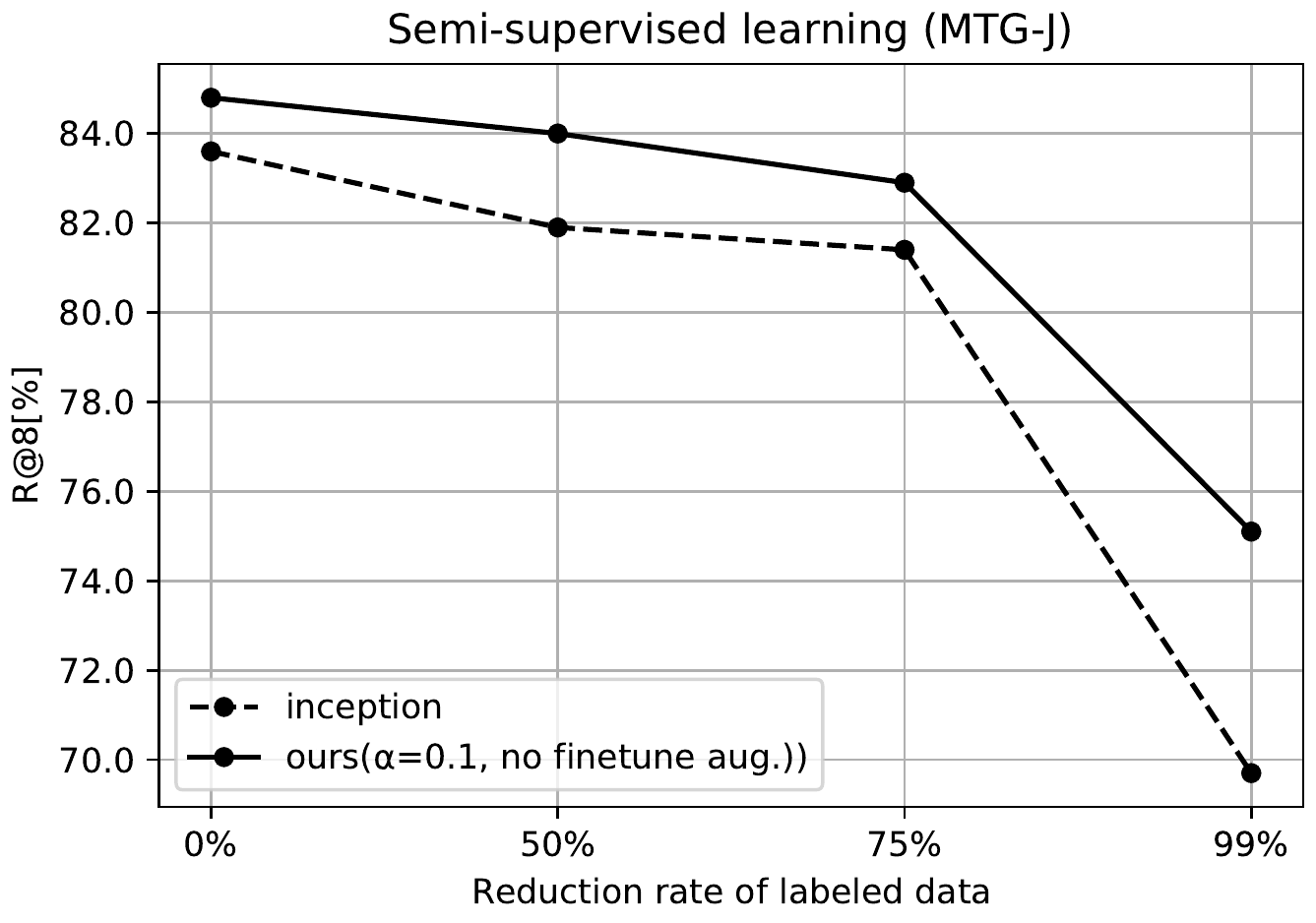}
  }
  \subfloat[ROC]{
  \includegraphics[width=0.32\textwidth, trim={0cm 0cm 0cm 0cm}, clip]{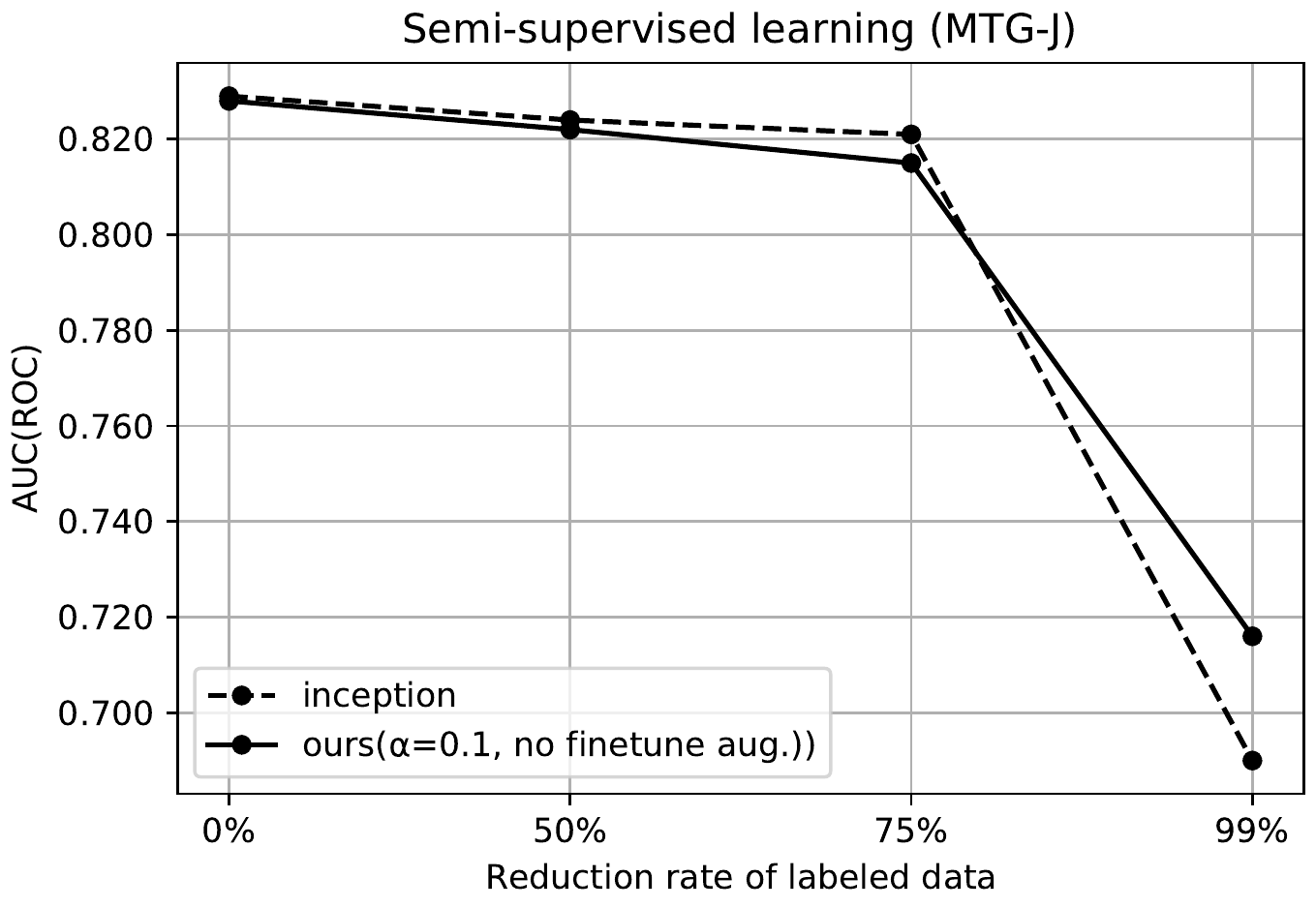}
  \label{fig:obj_eval}  }
 \subfloat[PR]{
    \includegraphics[width=0.32\textwidth, trim={0cm 0cm 0cm 0cm}, clip]{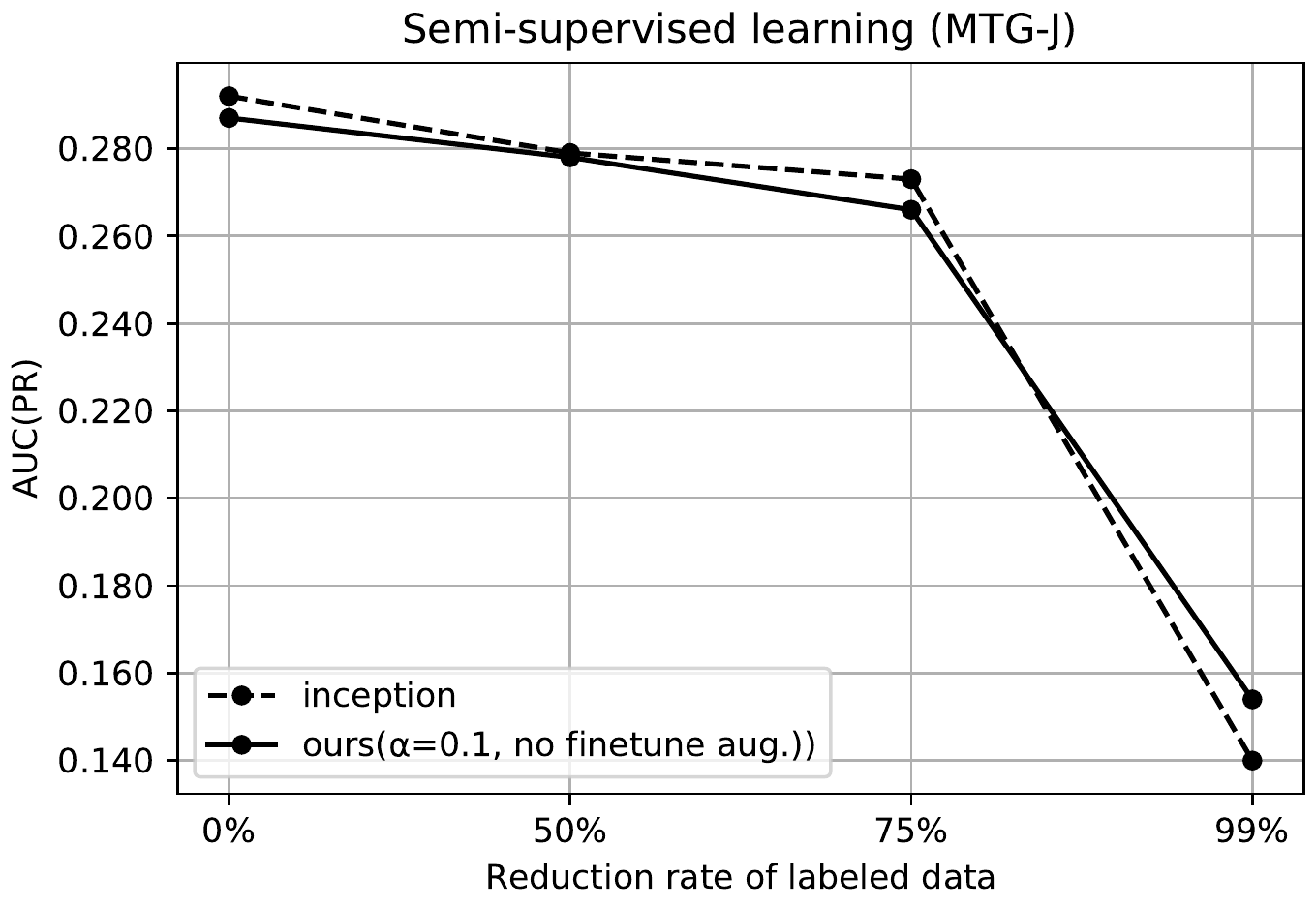}
  \label{fig:human_eval}
  }
  \caption{{\bf Results for semi-supervised scenario of MTG-Jamendo dataset.} (a-d) Similarity-based retrieval R@K results. (e-f) Auto-tagging AUC results.}
     \label{fig:mtg}
\end{figure*}

 Fig~\ref{fig:mtg} shows the results for the semi-supervised scenario of MTG-Jamendo dataset.
Similar to the MagnaTagATune dataset, compared to the inception model, the performance gain of our model tends to become larger as the amount of labeled data decreases.

\section{Related Work}\label{sec:relatedwork}
Spijkervet and Burgoyne demonstrated the effectiveness of SimCLR-based self-supervised learning for music auto-tagging \cite{CLMR}. We have shown that self-supervised learning is effective not only for auto-tagging but also for similarity-based music retrieval. Furthermore, our aim is to improve practical performance rather than merely evaluating representation quality. To this end, we propose a self-supervised auxiliary loss accompanied by a simple modified procedure that outperforms their self-supervised approach.

Thom{\'{e}} et al. introduced four triplet learning terms for learning music similarity, which include transformed excerpts, excerpts from the same track, and genre and mood membership \cite{MusicalSimilarity}. In contrast, our model employs SimCLR-based contrastive learning for self-supervised learning, manages general multi-tag settings through classification-based metric learning, addresses the auto-tagging task, and demonstrates effectiveness in semi-supervised settings.

Manocha et al. utilized SimCLR for pre-training, trained a loss net on JND data, and employed triplet comparison for learning \cite{CDPAM}. Their method focuses on speech similarity using carefully designed speech domain datasets, differing from our approach that targets global audio similarity in the music domain by leveraging widely available tag annotations.

Duan et al. employed self-supervised learning to train a teacher network \cite{SLADE}. Subsequently, they used the teacher network to generate pseudo labels, which were then utilized for metric learning with ranking loss. Our method applies self-supervision directly to the ``student'' network, eliminating the need for a teacher network. Additionally, their approach is designed for the image domain rather than music.

Fu et al. introduced an intra-class ranking loss in a self-supervised manner, in addition to metric learning for handling inter-class variance \cite{MetricSelfRanking}. However, their self-supervision employs intra-class ranking loss, which is distinct from our contrastive self-supervised loss, and their method is tailored to the image domain rather than music.

\section{Conclusion}\label{sec:conclusion}
In this paper, we presented a model that enhances the quality of music similarity-based retrieval and music auto-tagging. We explored the role of self-supervision in metric learning and proposed utilizing self-supervision as auxiliary loss for metric learning. Our model outperforms baseline methods and proves effective when human-provided music tags are limited. The music industry often deals with heterogeneous and extensive music databases characterized by long-tailed attributes. Human-annotated tags may be unavailable, unclean, or inconsistent across different database segments. We expect our approach, which generates learning signals without human annotation, to be effective in such real-world situations.

In principle, our methodology can be extended to other signal data types, such as bio-signals (EEG, ECG, EMG, etc.) and scientific measurements. We intend to apply our approach to bio-signals and investigate cross-modal retrieval as a means of bridging bio-signals and music.

\section{Data Availability}
The datasets generated and/or analysed during the current study are available in the sota-music-tagging-models repository, \url{https://github.com/minzwon/sota-music-tagging-models/tree/master/split/mtat}.

The datasets generated and/or analysed during the current study are available in the mtg-jamendo-dataset repository, \url{https://github.com/MTG/mtg-jamendo-dataset}.
  
\bibliographystyle{plain}
\bibliography{bib}

\begin{thebibliography}{10}

\bibitem{LayerNorm}
Lei~Jimmy Ba, Jamie~Ryan Kiros, and Geoffrey~E. Hinton.
\newblock Layer normalization.
\newblock {\em CoRR}, abs/1607.06450, 2016.

\bibitem{bogdanov2019mtg}
Dmitry Bogdanov, Minz Won, Philip Tovstogan, Alastair Porter, and Xavier Serra.
\newblock The mtg-jamendo dataset for automatic music tagging.
\newblock In {\em Machine Learning for Music Discovery Workshop, International
  Conference on Machine Learning (ICML 2019)}, 2019.

\bibitem{MusicalSimilarity}
Sebastian~Piwell Carl~Thom{\'{e}} and Oscar Utterb{\"{a}}ck.
\newblock Musical audio similarity with self-supervised convolutional neural
  networks.
\newblock {\em CoRR}, abs/2202.02112, 2022.

\bibitem{SimCLR}
Ting Chen, Simon Kornblith, Mohammad Norouzi, and Geoffrey~E. Hinton.
\newblock A simple framework for contrastive learning of visual
  representations.
\newblock In {\em Proceedings of the 37th International Conference on Machine
  Learning, {ICML} 2020, 13-18 July 2020, Virtual Event}, volume 119 of {\em
  Proceedings of Machine Learning Research}, pages 1597--1607. {PMLR}, 2020.

\bibitem{SLADE}
Jiali Duan, Yen{-}Liang Lin, Son~Dinh Tran, Larry~S. Davis, and C.{-}C.~Jay
  Kuo.
\newblock {SLADE:} {A} self-training framework for distance metric learning.
\newblock In {\em {CVPR}}, pages 9644--9653. Computer Vision Foundation /
  {IEEE}, 2021.

\bibitem{MetricSelfRanking}
Zheren Fu, Yan Li, Zhendong Mao, Quan Wang, and Yongdong Zhang.
\newblock Deep metric learning with self-supervised ranking.
\newblock In {\em {AAAI}}, pages 1370--1378. {AAAI} Press, 2021.

\bibitem{MagnaTagATune}
Edith Law, Kris West, Michael~I. Mandel, Mert Bay, and J.~Stephen Downie.
\newblock Evaluation of algorithms using games: The case of music tagging.
\newblock In {\em {ISMIR}}, pages 387--392. International Society for Music
  Information Retrieval, 2009.

\bibitem{MetricLearningISMIR2020}
Jongpil Lee, Nicholas~J. Bryan, Justin Salamon, Zeyu Jin, and Juhan Nam.
\newblock Metric learning vs classification for disentangled music
  representation learning.
\newblock In Julie Cumming, Jin~Ha Lee, Brian McFee, Markus Schedl, Johanna
  Devaney, Cory McKay, Eva Zangerle, and Timothy de~Reuse, editors, {\em
  Proceedings of the 21th International Society for Music Information Retrieval
  Conference, {ISMIR} 2020, Montreal, Canada, October 11-16, 2020}, pages
  439--445, 2020.

\bibitem{lee2018samplecnn}
Jongpil Lee, Jiyoung Park, Keunhyoung~Luke Kim, and Juhan Nam.
\newblock Samplecnn: End-to-end deep convolutional neural networks using very
  small filters for music classification.
\newblock {\em Applied Sciences}, 8(1):150, 2018.

\bibitem{CDPAM}
Pranay Manocha, Zeyu Jin, Richard Zhang, and Adam Finkelstein.
\newblock {CDPAM:} contrastive learning for perceptual audio similarity.
\newblock In {\em {IEEE} International Conference on Acoustics, Speech and
  Signal Processing, {ICASSP} 2021}.

\bibitem{SoftTripleLoss}
Qi~Qian, Lei Shang, Baigui Sun, Juhua Hu, Tacoma Tacoma, Hao Li, and Rong Jin.
\newblock Softtriple loss: Deep metric learning without triplet sampling.
\newblock In {\em {ICCV}}, pages 6449--6457. {IEEE}, 2019.

\bibitem{CLMR}
Janne Spijkervet and John~Ashley Burgoyne.
\newblock Contrastive learning of musical representations.
\newblock {\em CoRR}, abs/2103.09410, 2021.

\bibitem{MusicTagTrans}
Minz Won, Keunwoo Choi, and Xavier Serra.
\newblock Semi-supervised music tagging transformer.
\newblock In {\em {ISMIR}}, pages 769--776, 2021.

\bibitem{EvalMusicTag}
Minz Won, Andres Ferraro, Dmitry Bogdanov, and Xavier Serra.
\newblock Evaluation of cnn-based automatic music tagging models.
\newblock {\em CoRR}, abs/2006.00751, 2020.

\bibitem{ClassificationMetricLearning}
Andrew Zhai and Hao{-}Yu Wu.
\newblock Classification is a strong baseline for deep metric learning.
\newblock In {\em {BMVC}}, page~91. {BMVA} Press, 2019.

\end{thebibliography}

\end{document}